\documentclass[iop]{emulateapj}

\usepackage{apjfonts}
\newcommand{\ppcf}{Plasma Phys. Control. Fusion}
\newcommand{\lrsp}{Living Rev. Sol. Phys.}
\newcommand{\rslpsa}{Proc. R. Soc. A}
\newcommand{\aipcp}{AIP Conf. Proc.}
\newcommand{\pop}{Phys. Plasmas}
\newcommand{\Alfven}{Alfv\'{e}n}
\newcommand{\Alfvenic}{Alfv\'{e}nic}
\newcommand{\para}{\parallel}
\newcommand{\Esw}{\ensuremath{\mathbf{E}_{\text{sw}}}}
\newcommand{\Esc}{\ensuremath{\mathbf{E}_{\text{sc}}}}
\newcommand{\Eysw}{\ensuremath{E_{y,\text{sw}}}}
\newcommand{\Exsw}{\ensuremath{E_{x,\text{sw}}}}
\newcommand{\Eysc}{\ensuremath{E_{y,\text{sc}}}}
\newcommand{\Exsc}{\ensuremath{E_{x,\text{sc}}}}
\newcommand{\vsw}{\ensuremath{\mathbf{v}_{\text{sw}}}}

\shorttitle{SOLAR WIND ELECTRIC FIELD SPECTRUM}
\shortauthors{CHEN ET AL.}
\slugcomment{}
\def\andname{}

\begin{document}
\title{Frame Dependence of the Electric Field Spectrum of Solar Wind Turbulence}
\author{C.~H.~K.~Chen\altaffilmark{1}, S.~D.~Bale\altaffilmark{1,2}, C.~Salem\altaffilmark{1}, and F.~S.~Mozer\altaffilmark{1}}
\affil{$^1$Space Sciences Laboratory, University of California, Berkeley, CA 94720, USA; chen@ssl.berkeley.edu}
\affil{$^2$Physics Department, University of California, Berkeley, CA 94720, USA}
\begin{abstract}
We present the first survey of electric field data using the \emph{ARTEMIS} spacecraft in the solar wind to study inertial range turbulence. It was found that the average perpendicular spectral index of the electric field depends on the frame of measurement. In the spacecraft frame it is $-5/3$, which matches the magnetic field due to the large solar wind speed in Lorentz transformation. In the mean solar wind frame, the electric field is primarily due to the perpendicular velocity fluctuations and has a spectral index slightly shallower than $-3/2$, which is close to the scaling of the velocity. These results are an independent confirmation of the difference in scaling between the velocity and magnetic field, which is not currently well understood. The spectral index of the compressive fluctuations was also measured and found to be close to $-5/3$, suggesting that they are not only passive to the velocity but may also interact nonlinearly with the magnetic field.
\end{abstract}

\section{Introduction}

The solar wind is a plasma that is observed to be turbulent with fluctuations at a broad range of scales \citep{tu95,goldstein95,horbury05,bruno05a,petrosyan10}. It is usually modeled as a cascade of energy from large scales \citep[e.g.,][]{wicks10a}, where the energy is injected, to small scales \citep[e.g.,][]{chen10b}, where kinetic processes dissipate the energy \citep[e.g.,][]{schekochihin09}. The inertial range fluctuations are thought to be primarily  \Alfvenic\ in nature, with \Alfven-wave-like polarizations \citep{belcher71} and phase speeds close to the \Alfven\ speed \citep{bale05}.

There are various theories of \Alfvenic\ turbulence, based on interacting packets of \Alfven\ waves. The theory of \citet{goldreich95}, based on critical balance, predicts that the \Alfvenic\ fluctuations have a perpendicular one-dimensional energy spectrum $E(k_\perp)\sim k_\perp^{-5/3}$, where $k_\perp$ is the wavevector perpendicular to the magnetic field. The theory of \citet{boldyrev06}, which in addition assumes scale-dependent alignment, predicts that their spectrum is $E(k_\perp)\sim k_\perp^{-3/2}$. Similar predictions also exist for the multitude of imbalanced theories \citep[e.g.,][]{lithwick07,beresnyak08,chandran08,perez09,podesta10c,podesta11b}.

In the solar wind at 1 AU, it has been shown that the spectral index of the magnetic field is close to $-5/3$ on average but that the spectral index of the velocity is closer to $-3/2$ \citep[e.g.,][]{mangeney01,podesta07,salem09,tessein09,podesta10d,wicks11}. This difference between the two fields is not consistent with any of the current theories of \Alfvenic\ turbulence and is one of the currently unsolved problems of solar wind turbulence.

Past measurements of the electric field spectrum in the frame of the spacecraft found it to closely match the magnetic field \citep{bale05,sahraoui09}. These measurements used single intervals of data but it has been shown \citep[e.g.,][]{tessein09} that the velocity and magnetic field have a large spread of spectral indices and many intervals are needed to determine the average behavior.

In this Letter, we present a survey of electric field measurements in the solar wind using many intervals of data. We explain why the electric field in the spacecraft frame follows the magnetic field and make new measurements of the electric field in the mean solar wind frame. In Section \ref{sec:data}, we describe the data set, in Section \ref{sec:results} we discuss our results and in Section \ref{sec:conclusions} we present our conclusions.

\section{Data Set}
\label{sec:data}

We used data from the \emph{ARTEMIS} mission \citep{angelopoulos10}, which is an extension of the \emph{THEMIS} mission \citep{angelopoulos08}. During late 2010, the two \emph{ARTEMIS} spacecraft (\emph{P1} and \emph{P2}) moved from equatorial Earth orbits to Lunar Lagrange orbits ($\sim$ 60 $R_{\text{E}}$ from the Earth). Periods of solar wind data were selected in which each spacecraft was upstream of the Moon, out of Earth's ion foreshock and the required instruments were operational. The selected days are: days 245--257, 308--310, 316--318, 337--343 of 2010 and days 1--3, 40--42 of 2011 for \emph{P1}; days 217--230, 275--284, 304--307, 361--364 of 2010 and days 25--28 of 2011 for \emph{P2}. The same day in both spacecraft was avoided so that the intervals are independent. All of the data from these days were split into 6 hr sections resulting in 272 intervals, 98\% of which were in slow solar wind ($<$500 km s$^{-1}$).

Spin resolution ($\sim$3 s) electric field data, \Esc, from the electric field instrument \citep[EFI;][]{bonnell08} was used, along with spin resolution magnetic field data, $\mathbf{B}$, from the fluxgate magnetometer \citep[FGM;][]{auster08} and varying resolution ion velocity, $\mathbf{v}$, and ion number density, $n$, onboard moments from the electrostatic analyzer \citep[ESA;][]{mcfadden08}. A despun spacecraft coordinate system (DSL) was used, in which $z$ is the spacecraft spin axis. The DSL system for \emph{ARTEMIS} is approximately the same as the geocentric solar ecliptic (GSE) system with the sign of the $y$- and $z$- axes reversed. The wire boom electric field antennas are in the $x$--$y$ plane and extend a few Debye lengths from the spacecraft. Data with the currently most recent calibrations (v01) were used for all instruments.

For \Esc\ it was found that some extra calibration was needed. A least-squares fit, varying the \Eysc\ offset $O_{E_y}$ and the \Esc\ scaling factor $F$, was performed to minimize the difference between \Eysc\ and the $y$-component of $-\mathbf{v}\times\mathbf{B}$ for each interval (this technique assumes ideal MHD). Each 6 hr interval was corrected using these empirically determined values. The mean value of $O_{E_y}$ was found to be $-0.17$ mV m$^{-1}$ for \emph{P1} and $-0.23$ mV m$^{-1}$ for \emph{P2}; the mean value of $F$ was found to be 1.02 for \emph{P1} and 0.99 for \emph{P2}. An alternative fit using a $B_z$ offset instead of an \Eysc\ offset was also tried, resulting in $B_z$ offsets $\sim$ 0.6 nT. The results of this Letter, however, are not significantly affected by either of these additional calibration methods.

The velocity and density data was cleaned up by removing unphysical spikes and other spurious data that was present. This was done by linearly interpolating over data points more than 4 standard deviations from the mean in each 6 hr interval (this process was repeated three times for each interval). Any data gaps in the 3 s resolution data were also linearly interpolated over to produce time series with consistent 3 s resolution. After this process, occasional small spikes, sometimes seen in all three instruments, remained in the time series of some intervals. These are likely due to noise but did not affect this analysis (excluding these intervals did not significantly change the results of this Letter).

The electric field was measured in the frame of the spacecraft, \Esc, and converted into the frame of the mean solar wind velocity, \Esw, using the Lorentz transformation,
\begin{equation}
\label{eq:lorentz}
\mathbf{E}_{\text{sw}}=\mathbf{E}_{\text{sc}}+\mathbf{v}_{\text{sw}}\times\mathbf{B},
\end{equation}
where $\mathbf{v}_{\text{sw}}$ is the mean solar wind velocity relative to the spacecraft over each interval ($\mathbf{v}=\vsw+\delta\mathbf{v}$). Since $\mathbf{B}$ is a fluctuating quantity, it was linearly interpolated onto the times of \Esc\ so that the transformation could be done for each electric field measurement.

\begin{figure}
\epsscale{1.2}
\plotone{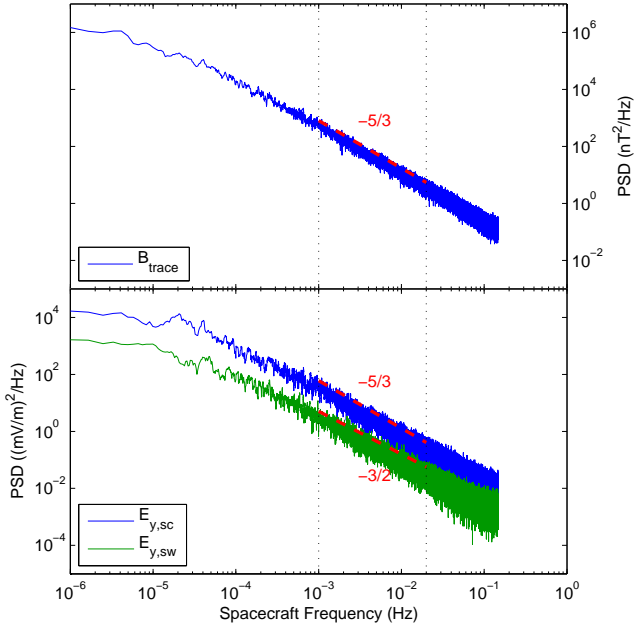}
\caption{\label{fig:spectra}Sample power spectra from \emph{P2} 2010 days 217--230. The dotted lines show the range of scales that the spectral index is fitted to. Gradients of $-5/3$ and $-3/2$ are marked for reference.}
\end{figure}

The power spectrum of each component of $\mathbf{B}$ and $\mathbf{v}$, of the $x$- and $y$- components of \Esc\ and \Esw, of the magnetic field magnitude $|\mathbf{B}|$, and of $n$ was calculated for each of the intervals. The multitaper method with time-bandwidth product $NW=4$ and 7 eigentapers \citep{percival93} was used (using a standard Fourier transform does not affect the results to within errors). Typical power spectra for a longer interval (14 days) are shown in Figure \ref{fig:spectra}: the trace of the magnetic field spectrum and the $y$-component of the electric field spectrum in both frames. Since the solar wind fluctuations are anisotropic with $k_\perp>k_\para$ \citep[e.g.,][]{chen10a}, these are measurements of the perpendicular spectrum $E(k_\perp)$ (to measure $E(k_\para)$ a local field tracking technique would be needed \citep[e.g.,][]{horbury08,chen11a}).

Each spectral index was determined from the gradient of the best-fit line to the power spectrum in log--log space over the spacecraft frequency range $1\times10^{-3}$ Hz to $2\times10^{-2}$ Hz (marked as dotted lines in Figure \ref{fig:spectra}). Applying Taylor's hypothesis \citep{taylor38} since the solar wind is super-\Alfvenic, this range corresponds approximately to scales 18,000 km to 350,000 km and a perpendicular wavevector range $0.0018<k_\perp\rho_i<0.036$, where $\rho_i\approx100$ km is the typical ion gyroradius. This is in the middle of the inertial range and was chosen because good power laws exist here in all intervals. The results of the analysis are described in the next section.

\section{Results}
\label{sec:results}

\begin{figure}
\epsscale{1.2}
\plotone{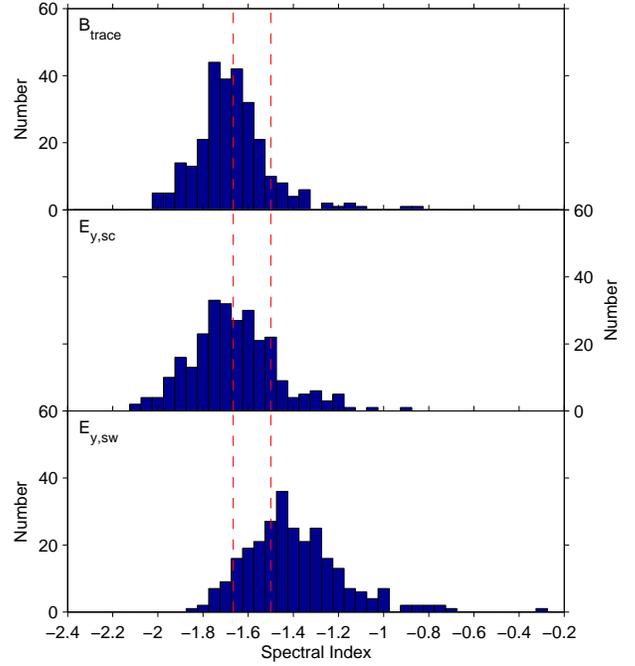}
\caption{\label{fig:histograms}Histograms of spectral index for the trace of the magnetic field and the $y$-component of the electric field in the spacecraft and mean solar wind frames. Values of $-5/3$ and $-3/2$ are marked for reference.}
\end{figure}

Histograms of the spectral indices for the magnetic and electric fields are shown in Figure \ref{fig:histograms}. The magnetic field trace spectral index histogram can be seen to peak close to $-5/3$, in agreement with previous results \citep[e.g.,][]{smith06a,tessein09}. The histogram of the $y$-component of the electric field in the spacecraft frame also peaks near $-5/3$ but the histogram of the same component in the mean solar wind frame peaks closer to $-3/2$.

The mean spectral indices for each field are given in Table~\ref{tab:scaling}, along with the standard error of the mean $\sigma/\sqrt{N}$, where $\sigma$ is the sample standard deviation and $N$ is the number of intervals. The mean velocity and density spectral indices were calculated from only 117 and 120 of the intervals, respectively. These are the intervals for which 3 s onboard moment data was available and no more than 5\% of the data was missing. This explains the larger error for these fields.  The same analysis was also tried with 24 hr intervals (not shown here), resulting in a smaller spread of spectral indices but the same mean values to within 2 standard errors.

\begin{deluxetable}{cc}
\tablecaption{\label{tab:scaling}Mean Spectral Indices}
\tablehead{\colhead{Field} & \colhead{Spectral Index}}
\startdata
$B_{\text{trace}}$ & $-1.67\pm 0.01$\\
$v_{\text{trace}}$ & $ -1.50\pm0.02$\\
\Eysc & $-1.66\pm 0.01$\\
\Exsc & $-1.45\pm 0.01$\\
\Eysw & $-1.40\pm 0.01$\\
\Exsw & $-1.39\pm 0.01$\\
$|\mathbf{B}|$ & $-1.64\pm0.01$\\
$n$ & $-1.63\pm0.02$
\enddata
\end{deluxetable}

The fact that the scaling of \Eysc\ matches $B_{\text{trace}}$ can be shown to be due to the Lorentz transformation. In ideal MHD, the three fields are related to each other in the mean solar wind frame by $\mathbf{E}_{\text{sw}}=-\delta\mathbf{v}\times\mathbf{B}$ and putting this into Equation (\ref{eq:lorentz}) gives
\begin{equation}
\label{eq:ideallorentz}
\mathbf{E}_{\text{sc}}=-\delta\mathbf{v}\times\mathbf{B}-\mathbf{v}_{\text{sw}}\times\mathbf{B}.
\end{equation}
Since the mean solar wind speed is much larger than the fluctuations, $|\mathbf{v}_{\text{sw}}|>|\delta\mathbf{v}|$, and is mostly in the $x$ (radial) direction, \Eysc\ is dominated by the magnetic field fluctuations convected by the mean solar wind flow and therefore follows their scaling. The amplitudes of the spectra in Figure \ref{fig:spectra} agree with this interpretation: the \Eysc\ spectrum is an order of magnitude larger than the \Eysw\ spectrum, showing that for the $y$-component, the second term on the right-hand side of Equation (\ref{eq:ideallorentz}) is larger than the first. The $x$-component of Eq.~\ref{eq:ideallorentz} does not depend on the radial solar wind velocity, so scaling of \Exsc\ does not depend only on the scaling of $\mathbf{B}$ and indeed is different to that of $B_{\text{trace}}$.

\Esw\ has a scaling closer to that of $v_{\text{trace}}$, which can also be shown to be consistent with \Alfvenic\ fluctuations in ideal MHD. Splitting the magnetic field into a constant mean value plus fluctuations, $\mathbf{B}=\mathbf{B}_0+\delta\mathbf{B}$, the electric field in the mean solar wind frame is given by
\begin{equation}
\mathbf{E}_{\text{sw}}=-\delta\mathbf{v}\times\left(\mathbf{B}_0+\delta\mathbf{B}\right).
\end{equation}
The mean value of $|\delta\mathbf{B}|/|\mathbf{B}_0|$ is between 0.1 and 0.4 for the range of scales to which spectral indices were fitted. Since at these small scales in the solar wind $\mathbf{B}_0>\delta\mathbf{B}$, the electric field spectrum in the mean solar wind frame is dominated by the velocity fluctuations, and therefore has a similar scaling. Similar arguments can be made for the \Alfvenic\ fluctuations in gyrokinetic theory \citep{schekochihin09}. The fact that we observe a spectral index close to $-3/2$ in \Exsw\ and \Eysw\ also suggests that the perpendicular velocity component has this scaling, which is in agreement with the results of \citet{chapman07}. The electric field scaling is also in agreement with previous measurements of the velocity trace spectral index \citep[e.g.,][]{tessein09,podesta10d}.

The scaling of the compressive fluctuations ($|\mathbf{B}|$ and $n$) is close to $-5/3$, matching the trace magnetic field spectrum, rather than the velocity spectrum. Previous observations \citep[e.g.,][]{marsch90b,bellamy05,issautier10} could not distinguish between $-5/3$ and $-3/2$ in the compressive fluctuations so this scaling is consistent with those observations. The compressive fluctuations are mainly due to the slow mode \citep{howes11b} and are sometimes thought to be passive to the \Alfvenic\ turbulence. Since their scaling matches the magnetic field, rather than the velocity, the nonlinearity cannot be due solely to passive convection and may include nonlinearities with the magnetic field. This supports the theories of compressible reduced MHD and kinetic reduced MHD \citep{schekochihin09}, in which the compressive fluctuations interact nonlinearly with both the magnetic field and velocity.

To test the significance of the difference between the mean spectral index values in Table~\ref{tab:scaling}, the $t$-test was applied. This is appropriate since the spectral indices appear to be normally distributed and are independent measurements. The $t$ value for differentiating between the spectral indices of $B_{\text{trace}}$ and \Eysc\ is $t=0.41$. This is smaller than the 95\% value of 1.96 for infinite degrees of freedom, showing that there is no statistically significant difference between the scaling of these two fields. For differentiating between the spectral indices of \Eysw\ and \Eysc\ the $t$ value is $t=15$, larger than the 95\% value, showing that these two fields have significantly different spectral indices. This confirms that the $-5/3$ and $-3/2$ difference is a statistically robust result.

To examine the cause of the spread of spectral index values, the correlation between the different spectral indices was measured. The linear correlation coefficients, calculated from various pairs of sets of the 272 spectral index values of each field, are shown in Table \ref{tab:correlations}. It can be seen that the spectral indices of most pairs of fields are poorly correlated, having correlation coefficients lower than 0.4. This suggests that the spread of values is mostly due to random, rather than systematic, variation, although the fact that the correlation coefficients are all slightly positive suggests perhaps some small underlying systematic variation. The exceptions are correlations between $B_{\text{trace}}$ and \Eysw\ and between \Exsc\ and \Exsw, which have correlation coefficients larger than 0.8. This is due to the reasons discussed above: the \Eysw\ spectrum is essentially a measure of the $B_{\text{trace}}$ spectrum because the $y$-component of the last term in Equation (\ref{eq:ideallorentz}) is large and \Exsc\ and \Exsw\ are similar because the $x$-component of the last term in Equation (\ref{eq:ideallorentz}) is not large (since \vsw\ is mostly in the $x$-direction).

\begin{deluxetable}{ccc}
\tablecaption{\label{tab:correlations}Correlations Between Spectral Indices}
\tablehead{\colhead{Field 1} & \colhead{Field 2} & \colhead{Correlation Coefficient}}
\startdata
$B_{\text{trace}}$ & \Eysc & 0.82\\
$B_{\text{trace}}$ & \Eysw & 0.15\\
$B_{\text{trace}}$ & \Exsc & 0.24\\
$B_{\text{trace}}$ & \Exsw & 0.08\\
\Eysc & \Eysw & 0.17\\
\Eysc & \Exsc & 0.24\\
\Eysc & \Exsw & 0.14\\
\Eysw & \Exsc & 0.38\\
\Eysw & \Exsw & 0.36\\
\Exsc & \Exsw & 0.86
\enddata
\end{deluxetable}

\section{Summary and Conclusions}
\label{sec:conclusions}

We have performed the first survey of electric field data in the solar wind to measure the perpendicular spectrum of inertial range fluctuations. It was found that there is a spread of spectral index values but that the average spectral index depends on the frame of measurement. In the spacecraft frame, the $y$-component of the electric field is primarily due to the magnetic field fluctuations being convected past the spacecraft at the average solar wind speed. It, therefore, has the same average spectral index (to within errors) as the magnetic field of $-1.66\pm0.01$. This is consistent with previous single interval electric field measurements in the spacecraft frame \citep{bale05,sahraoui09}. In the mean solar wind frame, the electric field is primarily due to velocity fluctuations in a mean magnetic field and has a spectral index of $-1.40\pm0.01$, which is closer to the velocity spectral index than the magnetic field spectral index, although not the same to within errors. The compressive fluctuations ($|\mathbf{B}|$ and $n$) were found to have the same spectral index as the magnetic field and not the velocity.

The difference between the scaling of the electric field in the spacecraft frame and the mean solar wind frame provides independent confirmation of the difference in scaling between the velocity and magnetic field. This difference is not expected for \Alfvenic\ fluctuations, since $\delta\mathbf{v}$ is proportional to $\delta\mathbf{B}$ in an \Alfven\ wave, and is not predicted by any of the current theories of \Alfvenic\ turbulence (although see recent work by \citet{boldyrev11} and \citet{wang11}). Recently,  \citet{roberts10} found that further out into the heliosphere, past 5 AU, the velocity spectral index evolves toward $-5/3$ to match the magnetic field. Although an important result, this does not explain the difference at 1 AU. Possible reasons for the difference include the effects of scale-dependent alignment, imbalance and  residual energy and these will be investigated in a future paper.

\acknowledgments
This work was supported by NASA grant NNX09AE41G. We acknowledge the \emph{THEMIS} team and NASA contract NAS5-02099. We thank J.~Bonnell, J.~McFadden, A.~Schekochihin, and D.~Sundkvist for useful conversations and an anonymous referee for helpful comments.

\end{document}